\documentclass[11pt]{article}
\usepackage[a4paper,margin=1in]{geometry}
\usepackage{amsmath,amssymb,amsthm,mathtools}
\usepackage[T1]{fontenc}
\usepackage[utf8]{inputenc}
\usepackage{lmodern}
\usepackage{microtype}
\usepackage{xcolor}
\usepackage{hyperref}

\title{Nelson’s Stochastic Mechanics:\\ Measurement, Nonlocality, and the Classical Limit}
\author{Partha Ghose\thanks{Email: partha.ghose@gmail.com}\\
Tagore Centre for Natural Sciences and Philosophy,\\
Rabindra Tirtha, New Town, Kolkata 700156, India}
\date{}

\begin{document}
\maketitle

\begin{abstract}
Nelson's stochastic mechanics may be understood as a stochastic underpinning, or reconstruction, of nonrelativistic quantum mechanics, once the diffusion scale is fixed by $\hbar$ and the admissible states are restricted by the usual single-valuedness condition on the wavefunction. In this note I briefly indicate what this route achieves and why it remains conceptually attractive. Four advantages are emphasized. First, it supplies a clear configuration-space stochastic picture of the underlying processes. Second, the Born rule is built in from the outset, with $|\psi|^2$
arising as the probability density $\rho$ of the underlying diffusion process rather than as an independent postulate. Third, it offers a markedly different perspective on measurement and nonlocality: in particular, collapse need not be treated as an extra axiom, and the nonlocality associated with entangled states is softened relative to the deterministic Bohmian guidance picture. Fourth, by tying quantumness to a diffusion scale, it naturally suggests a continuum of physical descriptions ranging from the strictly classical to the strictly quantum-mechanical regime. I conclude by proposing a natural distance scale in stochastic mechanics and examining its implications for testing possible limits of Bell correlations.
\end{abstract}

\section{Introduction}

The orthodox formulation of nonrelativistic quantum mechanics is extraordinarily successful, but it is not conceptually transparent in the same way as classical statistical mechanics. The wavefunction is postulated, the Born rule is added, and the measurement problem is then inherited in familiar form. Nelson's stochastic mechanics begins elsewhere. It starts from a configuration-space diffusion process and asks under what conditions the Schr\"odinger equation emerges from it. The resulting framework does not replace the empirical content of standard quantum mechanics; rather, it supplies a constructive route to it in probabilistic terms \cite{Nelson1966,Nelson1985,Bacciagaluppi2005}.

The point of the present note is modest. Compared to standard quantum mechanics and Bohmian mechanics, it has drawn much less attention. I argue that it deserves renewed attention because, once its exact achievement is stated carefully, it exhibits several advantages over the orthodox and Bohmian routes.

\section{What Nelson's route achieves}

In Nelson's framework one begins with a stochastic process $X(t)$ in configuration space, with forward and backward drifts. The sample paths are diffusion paths and are therefore almost surely nowhere differentiable; the theory accordingly replaces ordinary pathwise velocities by mean forward and backward conditional drifts

\[
b(x,t)=\lim_{\Delta t\to 0^+}\,
\mathbb{E}\!\left[
\frac{X(t+\Delta t)-X(t)}{\Delta t}
\,\middle|\, X(t)=x
\right],
\]
\[
b_{*}(x,t)=\lim_{\Delta t\to 0^+}\,
\mathbb{E}\!\left[
\frac{X(t)-X(t-\Delta t)}{\Delta t}
\,\middle|\, X(t)=x
\right].
\]
From these one defines the current and osmotic velocities,
\begin{equation}
  v = \frac{b+b_*}{2} = \frac{1}{m}\nabla S, \qquad u = \frac{b-b_*}{2} = \frac{\sigma}{2}\nabla \log \rho
\end{equation}
in terms of the action $S$, the probability density $\rho$ of the diffusion process and the diffusion coefficient $\sigma = \hbar/m$ per unit mass (equivalent, up to conventional factors, to Nelson's notation $\nu = \hbar/2m$).  Under the usual time-symmetric dynamical assumptions, one obtains the Madelung equations associated with the Schr\"odinger dynamics \cite{Nelson1966,Nelson1985}:
\[
\frac{\partial \rho}{\partial t}
+\nabla\!\cdot\!\left(\rho\,\frac{\nabla S}{m}\right)=0,
\]
\[
\frac{\partial S}{\partial t}
+\frac{(\nabla S)^2}{2m}
+V
-\frac{\hbar^2}{2m}\,
\frac{\nabla^2 \sqrt{\rho}}{\sqrt{\rho}}
=0.
\]
If the wavefunction is written in the form
\begin{equation}
  \psi = \sqrt{\rho}\,e^{iS/\hbar},\label{born}
\end{equation}
it satisfies the Schr\"{o}dinger equation
\[
i\hbar\,\frac{\partial \psi}{\partial t}
=
\left[
-\frac{\hbar^{2}}{2m}\nabla^{2}+V
\right]\psi .
\]
This yields the Schr\"odinger equation within the stochastic framework. If the admissible states are restricted by the usual single-valuedness condition on $\psi$ as in standard quantum mechanics, Nelson's framework reproduces the ordinary Schr\"odinger version of quantum mechanics \cite{Nelson1966,Nelson1985,Bacciagaluppi2005}.

An important advantage of Nelson's stochastic mechanics is that the \emph{Born rule} is not an independent postulate. The theory begins with a diffusion process \(X(t)\) in configuration space, whose probability density is \(\rho(x,t)\). When the Schr\"odinger equation is recovered within this framework, one finds (see eqn (\ref{born}))
\[
\rho(x,t)=|\psi(x,t)|^2 .
\]
The Born rule is therefore built into the theory at the foundational level: it expresses the distribution of the underlying stochastic process itself. This is conceptually preferable to the orthodox formulation, in which the rule \( |\psi|^2 \) must be added separately as a probabilistic postulate.

Thus, whereas orthodox quantum mechanics takes as foundational postulates the Schr\"odinger equation, the Born rule, and the single-valuedness of the wave function, Nelson's stochastic mechanics improves on it in two respects. First, the Schr\"odinger equation is not postulated but reconstructed from an underlying stochastic dynamics. Second, the Born rule is not an independent axiom, since \(|\psi|^2\) arises directly as the probability density of the diffusion process. Only one supplementary condition remains to be imposed, namely the usual single-valuedness of the wave function.
This is the first point that should be stated clearly: Nelson’s approach provides a stochastic underpinning, or reconstruction, of orthodox quantum mechanics while retaining the same empirical predictions.

\section{A clearer underlying picture}

The first conceptual advantage is that the theory provides a concrete stochastic picture of the underlying process. Standard quantum mechanics, taken by itself, yields the correct probabilities and transition amplitudes but does not provide a detailed spacetime account of what underlies them. Nelson's theory does. The primitive object is not an abstract state vector alone, but a diffusion in configuration space whose probability density evolves in such a way as to generate the quantum formalism \cite{Nelson1985,Bacciagaluppi2005}.

This point is important because it avoids a false choice between classical trajectories and sheer formalism. Nelson's process is not a classical trajectory in the Newtonian sense, since the paths are nondifferentiable. But it is still a perfectly definite stochastic process. Thus one gains a picture of physical evolution without being forced into the deterministic guidance structure of de Broglie--Bohm theory \cite{Bohm1952I,Bohm1952II}.

\section{Measurement and the softening of nonlocality}

A second advantage concerns measurement. In the orthodox framework, one typically supplements the unitary Schr\"odinger evolution by an additional collapse postulate or some functional equivalent. By contrast, within Nelson's framework Pavon showed that, at least for position measurements, the effective collapse of the wavefunction can be obtained from a purely probabilistic mechanism internal to the stochastic theory itself \cite{Pavon1999}. The essential point is that the update is no longer an alien axiom added to the dynamics from outside; it is treated as a change of the stochastic process under the relevant measurement constraint.

This does not merely shift notation. It changes the conceptual status of measurement. One need not posit a separate ``Process~1'' in the von Neumann sense. Measurement update can be understood as a conditioning and re-selection of the underlying stochastic dynamics. In that important sense, Nelson's approach substantially alleviates the traditional measurement problem \cite{Pavon1999,Bacciagaluppi2005}.

The same stochastic setting also leads to a more nuanced view of nonlocality. Einstein's concern was the apparent presence of ``spooky action at a distance'' in the EPR situation \cite{EPR1935}. Bohmian mechanics answers this concern in a very explicit way: it provides a deterministic guidance law in configuration space, and for entangled states the instantaneous velocity of one particle depends on the total configuration \cite{Bohm1952I,Bohm1952II}. Bell then showed that no theory satisfying the relevant locality requirements can reproduce all the quantum correlations \cite{Bell1964}.

Nelson's theory does not make Bell-type nonseparability disappear. The stochastic process is again defined in configuration space, and entangled states remain nonseparable in the relevant sense. But the dynamical realization of this fact is different. Since diffusion paths possess no sharp classical instantaneous velocity, the dependence is not expressed through a deterministic guidance condition of the Bohmian sort. It is encoded instead in the forward/backward drift structure, or equivalently in the current and osmotic velocities. The nonlocality is therefore not removed, but it is softened: what remains is configuration-space stochastic nonseparability rather than a sharp pathwise action-at-a-distance picture \cite{Nelson1985,Bacciagaluppi2005}.

This distinction matters. The Bell problem survives, but its ontological reading changes. Nelson's framework retains the nonseparability demanded by quantum correlations while avoiding the most rigid deterministic form of Bohmian nonlocality.

However, Bell’s theorem, especially in its Bohm–Aharonov spin form, does not turn on that distinction. For Bell-type spin correlations, Nelson offers no obvious advantage: the problem of local causality remains.

\section{A continuum from the classical to the quantum}

A third advantage is that Nelson's route naturally suggests a continuous family of physical descriptions between the classical and quantum domains. In the orthodox formulation, classical mechanics and quantum mechanics are often presented as conceptually disjoint, with the classical world brought in only as a limiting or external framework. In the stochastic approach the difference is tied instead to the diffusion scale. 

The classical endpoint itself can be written in wave-mechanical language, as Rosen showed long ago in his classical Schr\"odinger equation \cite{Rosen1964}. The conceptual lesson is that classical and quantum descriptions need not be separated by an absolute discontinuity. A continuous interpolation is physically thinkable.

Rosen's ``classical Schr\"odinger equation'' is obtained by subtracting the quantum
potential term from the standard Schr\"odinger equation, namely
\[
i\hbar\,\frac{\partial \psi}{\partial t}
=
\left[
-\frac{\hbar^{2}}{2m}\nabla^{2}+V-Q
\right]\psi 
\]
with
\[
Q=-\frac{\hbar^{2}}{2m}\frac{\nabla^{2}R}{R},
\]
the quantum potential. This is a nonlinear modification of the Schr\"{o}dinger equation. Its polar decomposition yields the classical Hamilton--Jacobi equation together with
the continuity equation:

\[
\frac{\partial S}{\partial t}
+\frac{(\nabla S)^2}{2m}
+V =0,
\]
\[
\frac{\partial \rho}{\partial t}
+\nabla\!\cdot\!\left(\rho\,\frac{\nabla S}{m}\right)=0,
\]
It is therefore natural to introduce a parameter
\(\lambda\) multiplying the quantum-potential contribution in the Rosen equation and write it in the form
\[
i\hbar\,\frac{\partial \psi}{\partial t}
=
\left[
-\frac{\hbar^{2}}{2m}\nabla^{2}+V-\lambda Q
\right]\psi .
\]
In polar form this gives
\[
\frac{\partial S}{\partial t}
+\frac{(\nabla S)^{2}}{2m}
+V
+(1-\lambda)Q
=0,
\]
together with the usual continuity equation. Thus \(\lambda=0\) reproduces standard
quantum mechanics, while \(\lambda=1\) yields the Rosen equation and hence the classical Hamilton-Jacobi equation. Intermediate values \(0<\lambda<1\) may be interpreted phenomenologically as coding the effect
of the environment, in the sense that environmental influence progressively suppresses
the quantum-potential term and drives the dynamics from the strictly quantum regime
towards the classical one. This is different from the quantum potential itself decreasing with larger mass $m$ or equivalently smaller $\sigma$ ($\hbar = m\sigma$) which are quantum mechanical effects.

A further clarification is useful at the classical endpoint. At \(\lambda=1\) the
Rosen equation yields the classical Hamilton--Jacobi equation together with the
continuity equation, but if one continues to write the state in the form
\(\psi=\sqrt{\rho}\,e^{iS/\hbar}\), a residual phase remains in the wave-mechanical
representation. For full classicality this residual phase should be regarded as
physically irrelevant and moded out. In other words, the classical limit is obtained
not merely by cancelling the quantum potential, but also by quotienting the remaining
\(U(1)\) phase freedom so that no interference significance is attached to it. This is
in keeping with the Koopman--von Neumann--Sudarshan viewpoint, in which classical
mechanics admits a Hilbert-space representation only after the physically irrelevant
phase is removed by a superselection rule or equivalent quotient construction
\cite{Koopman1931,vonNeumann1932,Sudarshan1976}.

This point is especially suggestive for foundational work. It opens the possibility that the so-called quantum--classical transition is not merely an issue of decoherence of already-given quantum states, but may reflect a more basic variation in the effective stochastic structure underlying the dynamics.

It is worth noting that the present viewpoint differs from Bohmian analyses of the classical Schrödinger equation. In Ghose \cite{Ghose2002} the modified equation was proposed not merely as a Bohmian rewriting, but as an alternative wave-mechanical formulation of classical dynamics together with an environment-dependent interpolation 
$\lambda$ linking the quantum and classical domains. By contrast, later Bohmian studies such as those of Benseny \emph{et al.} \cite{BensenyTenaOriols2016} and Navia--Sanz \cite{NaviaSanz2024} emphasize that the classical Schrödinger equation, when supplemented by a single-valued phase, still retains nonclassical features such as configuration-space noncrossing and residual coherence. The difference is therefore not merely technical: the earlier formulation assigns a direct physical meaning to the interpolating parameter and to the nonlinear term, whereas the later Bohmian analyses mainly explore the dynamical consequences and limitations of the resulting equation.

\section{A Possible Empirical Cutoff in Configuration Space?}
Because Nelson's theory assigns an explicit stochastic dynamics on configuration space, it is meaningful to ask whether this dynamics is valid on all scales or only up to some finite cutoff. Standard quantum mechanics contains no such intrinsic scale in its correlation structure: if entanglement is preserved, the correlations are in principle independent of spatial separation. A cutoff introduced into the Nelsonian dynamics would therefore lead to a directly testable distinction between the two theories. In that case, Nelson's theory would reproduce standard quantum mechanics only below the cutoff scale, while beyond it Bell-type and related nonlocal correlations should exhibit controlled deviations. The question of a cutoff is therefore not merely formal, but an empirical one.

\subsection*{A schematic cutoff scale}
To make the above possibility more explicit, one may introduce a phenomenological cutoff scale \(L_c\) into the Nelsonian stochastic dynamics and regard standard quantum mechanics as the limiting case \(L_c\to\infty\). Then the theory would reproduce the usual quantum correlations whenever the relevant induced separation scale \(L\) satisfies
\[
L \ll L_c,
\]
while deviations would be expected when
\[
L \gtrsim L_c.
\]
Schematically, one may write the correlation function in the form
\[
E(a,b;L)=E_{\mathrm{QM}}(a,b)\,F\!\left(\frac{L}{L_c}\right),
\]
where
\[
F(0)=1,
\qquad
F\!\left(\frac{L}{L_c}\right)\to 1
\quad \text{for} \quad L\ll L_c,
\]
and \(F(L/L_c)\) departs from unity as \(L\) becomes comparable with or larger than \(L_c\). In this way the question becomes directly empirical: one can test up to what scale Nelson's stochastic description remains indistinguishable from standard quantum mechanics, and whether a breakdown of exact quantum correlations occurs beyond that scale.

\section{Conclusion}

Nelson's stochastic mechanics should be appreciated for what it actually offers. It reconstructs nonrelativistic quantum mechanics from a configuration-space diffusion process once the diffusion scale is fixed by $\hbar$ and the usual single-valuedness condition is imposed on the wavefunction. So understood, it possesses three notable conceptual advantages.

First, it provides a clear stochastic picture of the underlying processes. Second, it changes the status of measurement and softens the interpretation of nonlocality: collapse can be treated within the stochastic framework itself, and entangled nonseparability need not be read in the hard deterministic Bohmian sense. Third, it offers a natural conceptual bridge between the classical and quantum regimes by tying quantumness to a diffusion scale.

It also suggests a concrete empirical question: whether the stochastic substructure remains equivalent to standard quantum mechanics at arbitrarily large separations, or only up to a finite Bell-correlation scale.

For these reasons Nelson's route deserves to be regarded not as a historical curiosity, but as a serious alternative starting point for thinking about quantum theory.

\section{Acknowledgements}
I acknowledge use of ChatGPT for checking and language polishing.


\begin{thebibliography}{99}

\bibitem{Nelson1966}
E.~Nelson,
\newblock ``Derivation of the Schr\"odinger equation from Newtonian mechanics,''
\newblock \emph{Physical Review} \textbf{150} (1966), 1079--1085.
\newblock DOI: \href{https://doi.org/10.1103/PhysRev.150.1079}{10.1103/PhysRev.150.1079}.
\newblock Open PDF: \href{https://www.physics.utah.edu/~lebohec/ScaleRelativity/Papers/1966_ENelson_Derivation_of_SchrodEq_from_NewtMech.pdf}{download}.

\bibitem{Nelson1985}
E.~Nelson,
\newblock \emph{Quantum Fluctuations},
\newblock Princeton University Press, Princeton, 1985.
\newblock Open PDF: \href{https://web.math.princeton.edu/~nelson/books/qf.pdf}{download}.

\bibitem{Bacciagaluppi2005}
G.~Bacciagaluppi,
\newblock ``A conceptual introduction to Nelson's mechanics,'' in
\emph{Quantum Mechanics: Are There Quantum Jumps? and On the Present Status of Quantum Mechanics},
A.~Bassi, D.~D\"urr, T.~Weber, N.~Zangh\`i (eds.), World Scientific, Singapore, 2005, pp.~367--388.
\newblock Open PDF: \href{https://philsci-archive.pitt.edu/8853/1/Nelson-revised.pdf}{download}.

\bibitem{Pavon1999}
M.~Pavon,
\newblock ``Derivation of the wave function collapse in the context of Nelson's stochastic mechanics,''
\newblock \emph{Journal of Mathematical Physics} \textbf{40} (1999), 5565--5577.
\newblock Article page: \href{https://pubs.aip.org/aip/jmp/article/40/11/5565/398780/Derivation-of-the-wave-function-collapse-in-the}{AIP}.
\newblock Preprint/Open access: \href{https://arxiv.org/abs/quant-ph/9912015}{arXiv}.

\bibitem{EPR1935}
A.~Einstein, B.~Podolsky, and N.~Rosen,
\newblock ``Can quantum-mechanical description of physical reality be considered complete?''
\newblock \emph{Physical Review} \textbf{47} (1935), 777--780.
\newblock DOI: \href{https://doi.org/10.1103/PhysRev.47.777}{10.1103/PhysRev.47.777}.
\newblock Open PDF: \href{https://cds.cern.ch/record/405662/files/PhysRev.47.777.pdf}{download}.

\bibitem{Bohm1952I}
D.~Bohm,
\newblock ``A suggested interpretation of the quantum theory in terms of `hidden' variables. I,''
\newblock \emph{Physical Review} \textbf{85} (1952), 166--179.
\newblock DOI: \href{https://doi.org/10.1103/PhysRev.85.166}{10.1103/PhysRev.85.166}.
\newblock PDF mirror: \href{https://promptrevolution.poltextlab.com/content/files/2025/03/Bohm_1952.pdf}{download}.

\bibitem{Bohm1952II}
D.~Bohm,
\newblock ``A suggested interpretation of the quantum theory in terms of `hidden' variables. II,''
\newblock \emph{Physical Review} \textbf{85} (1952), 180--193.
\newblock DOI: \href{https://doi.org/10.1103/PhysRev.85.180}{10.1103/PhysRev.85.180}.
\newblock Open PDF: \href{https://cqi.inf.usi.ch/qic/bohm2.pdf}{download}.

\bibitem{Bell1964}
J.~S.~Bell,
\newblock ``On the Einstein Podolsky Rosen paradox,''
\newblock \emph{Physics} \textbf{1} (1964), 195--200.
\newblock DOI: \href{https://doi.org/10.1103/PhysicsPhysiqueFizika.1.195}{10.1103/PhysicsPhysiqueFizika.1.195}.
\newblock Open PDF: \href{https://cds.cern.ch/record/111654/files/vol1p195-200_001.pdf}{download}.

\bibitem{Rosen1964}
N.~Rosen,
\newblock ``The relation between classical and quantum mechanics,''
\newblock \emph{American Journal of Physics} \textbf{32} (1964), 597--600.


\bibitem{Koopman1931}
B.~O.~Koopman,
\newblock ``Hamiltonian Systems and Transformation in Hilbert Space,''
\newblock \emph{Proceedings of the National Academy of Sciences of the USA} \textbf{17}(5) (1931) 315--318.
\newblock DOI: \href{https://doi.org/10.1073/pnas.17.5.315}{10.1073/pnas.17.5.315}
.
\newblock Downloadable PDF: \href{https://www.pnas.org/doi/pdf/10.1073/pnas.17.5.315}{PNAS
 PDF}.

\bibitem{vonNeumann1932}
J.~von~Neumann,
\newblock ``Zur Operatorenmethode in der klassischen Mechanik,''
\newblock \emph{Annals of Mathematics} \textbf{33}(3) (1932) 587--642.
\newblock DOI: \href{https://doi.org/10.2307/1968537}{10.2307/1968537}
.

\bibitem{Sudarshan1976}
E.~C.~G.~Sudarshan,
\newblock ``Interaction between Classical and Quantum Systems and the Measurement of Quantum Observables,''
\newblock \emph{Pramana -- Journal of Physics} \textbf{6}(3) (1976) 117--126.
\newblock DOI: \href{https://doi.org/10.1007/BF02847120}{10.1007/BF02847120}
.
\newblock Downloadable PDF: \href{https://web2.ph.utexas.edu/~gsudama/pub/1976_005.pdf}{author
 PDF}.

\bibitem{Ghose2002}
P.~Ghose,
\newblock ``A Continuous Transition Between Quantum and Classical Mechanics. I,''
\newblock \emph{Foundations of Physics} \textbf{32}(6) (2002) 871--892.
\newblock DOI: \href{https://doi.org/10.1023/A:1016055128428}{10.1023/A:1016055128428}
.
\newblock Preprint: \href{https://arxiv.org/abs/quant-ph/0104104}{arXiv:quant-ph/0104104}
.

\bibitem{BensenyTenaOriols2016}
A.~Benseny, D.~Tena, and X.~Oriols,
\newblock ``On the Classical Schr\"{o}dinger Equation,''
\newblock \emph{Fluctuation and Noise Letters} \textbf{15}(3) (2016) 1640011.
\newblock DOI: \href{https://doi.org/10.1142/S0219477516400113}{10.1142/S0219477516400113}
.
\newblock Preprint: \href{https://arxiv.org/abs/1607.00168}{arXiv:1607.00168}
.

\bibitem{NaviaSanz2024}
D.~Navia and \'A.~S.~Sanz,
\newblock ``Exploring the nonclassical dynamics of the `classical' Schr\"{o}dinger equation,''
\newblock \emph{Annals of Physics} \textbf{463} (2024) 169637.
\newblock DOI: \href{https://doi.org/10.1016/j.aop.2024.169637}{10.1016/j.aop.2024.169637}
.
\newblock Preprint: \href{https://arxiv.org/abs/2312.02977}{arXiv:2312.02977}.


\end{thebibliography}
\end{document}